\documentclass[ 
                          showpacs, 
                          nofootinbib, 
                          nobibnotes, 
                          aps,  
                          superscriptaddress, 
                          amssymb, 
                          amsmath, 
                          floatfix, 
                          reprint, 
                          pra, 
                          notitlepage, 
                          showkeys, 
                          10pt]{revtex4-1}

\usepackage{amssymb}
\usepackage{amsmath}
\usepackage{graphicx}
\usepackage{dcolumn}
\usepackage[colorlinks=true,breaklinks=true,allcolors=blue]{hyperref}
\usepackage{url}
\usepackage{todonotes}
\usepackage{tikz}
\usepackage{dsfont}
\usepackage{epsfig}
\usepackage{xcolor}
\usepackage{bbold}
\usepackage{cancel}

\usetikzlibrary{decorations.text}

\usetikzlibrary{shapes.symbols, shapes.arrows, calc, shapes}

\def\equationautorefname~#1\null{Equation~(#1)\null}

\DeclareMathOperator{\Trace}{Tr}
\newcommand{\Tr}[1]{\Trace\left[{#1}\right]}

\renewcommand{\ln}[1]{\mathrm{ln}\left({#1}\right)}

\newcommand{\Eq}[1]{Eq.\,\eqref{eq:#1}}

\newcommand{\Fig}[1]{Fig.~\ref{fig:#1}}
\newcommand{\fig}[1]{\ref{fig:#1}}

\newcommand{\1}{\uparrow}

\newcommand{\corg}[1]{{\color{orange}{((#1))}}}
\renewcommand{\corg}[1]{{\color{orange}{}}}
\newcommand{\corgTG}[1]{{\color{orange}{((TG: #1))}}}
\renewcommand{\corgTG}[1]{{\color{orange}{}}}
\newcommand{\corgMK}[1]{{\color{red}{(({\em MK: #1}))}}}
\renewcommand{\corgMK}[1]{{\color{red}{}}}

\makeatletter
\let\cat@comma@active\@empty
\makeatother

\begin{document}

\title{Quenches near criticality of the quantum Ising chain---power and limitations of the\\ discrete truncated Wigner approximation}

\author{Stefanie Czischek}
\author{Martin G\"arttner}
\author{Markus Oberthaler}
\affiliation{Kirchhoff-Institut f\"ur Physik, Ruprecht-Karls-Universit\"at Heidelberg, Im Neuenheimer Feld 227, 69120 Heidelberg, Germany}
\author{Michael Kastner}
\affiliation{National Institute for Theoretical Physics (NITheP), Stellenbosch 7600, South Africa}
\affiliation{Department of Physics, Institute of Theoretical Physics, University of Stellenbosch, Stellenbosch 7600, South Africa}
\author{Thomas Gasenzer}
\affiliation{Kirchhoff-Institut f\"ur Physik, Ruprecht-Karls-Universit\"at Heidelberg, Im Neuenheimer Feld 227, 69120 Heidelberg, Germany}
\date{\today}

\begin{abstract}
The semi-classical discrete truncated Wigner approximation (dTWA) has recently been proposed as a simulation method for spin-$1/2$ systems. 
While it appears to provide a powerful approach which shows promising results in higher dimensions and for systems with long-range interactions, its performance is still not well understood in general. 
Here we perform a systematic benchmark on the one-dimensional transverse-field Ising model and point to limitations of the approximation arising after sudden quenches into the quantum critical regime. 
Our procedure allows to identify the limitations of the semi-classical simulations and with that to determine the regimes and questions where quantum simulators can provide information which is inaccessible to semi-classics.
\end{abstract}

\maketitle

\section{Introduction}
The efficient numerical simulation of dynamics in spin systems out of equilibrium is still an outstanding problem, as the exponentially growing Hilbert space limits calculations on classical computers.
For spin chains in one spatial dimension, excellent simulation methods based on matrix product states (MPS), such as the time-dependent density-matrix renormalisation group (tDMRG) approach, are available  \cite{White1992,Schollwoeck2011, Vidal2004a, Daley2004a, Sharma2015, Haegeman2016}. 
While these methods can give numerically exact results, they are limited to one-dimensional systems and break down in regimes of large entanglement entropy, e.g.~quantum critical regimes. There, the necessary computational resources, again, scale exponentially with system size, rendering the method eventually inefficient.

In such quantum critical regimes, efficient simulation methods are scarce. 
Recently, the discrete truncated Wigner approximation (dTWA) \cite{Wootters1987} has been proposed as a semi-classical simulation approach for spin-$1/2$ systems \cite{Schachenmayer2015, Schachenmayer2015b}. 
As a discrete variant of the well established truncated Wigner approximation for continuous systems, the dTWA is based on sampling the initial states from a Wigner function on a discrete phase space, and time-evolving each of the sample points according to semi-classically approximated equations of motion.
The dTWA, in particular, is not limited to one-dimensional systems.
Furthermore, while it is expected to develop its maximum power at short times, it has been found to accurately capture the dynamics in strongly correlated systems for longer evolution times \cite{Pucci2016}.

Because of these features, the method seems to open a promising route to simulating discrete models in regimes and dimensions where other simulation methods like tDMRG break down. 
dTWA has been used for simulating Heisenberg spins \cite{Babadi2015}, spin-boson models \cite{Orioli2017}, Ryd\-berg systems \cite{Orioli2018}, lattices of dipolar molecules \cite{Covey2017}, as well as for investigating many-body localisation and thermalisation \cite{Acevedo2017}. 
For increasing its accuracy, extensions and refinements of the method have been proposed \cite{Zunkovic2015, Pucci2017, Cai2018b}. 
Nevertheless, it is not clear yet under which circumstances the dTWA shows accurate results and which limitations it is subject to.

Here, we provide a benchmark of the dTWA, including regimes in which strong correlations prevail.
We analyse, in particular, the method's performance in describing the out-of-equilibrium dynamics of the one-dimensional transverse-field Ising model (TFIM) when quenched into the vicinity of its quantum critical point. 
The TFIM can be solved analytically and shows a quantum phase transition at zero temperature as a  function of the strength of the transverse magnetic field \cite{Pfeuty1970a,Calabrese2012a,Calabrese2012b}. 
Our comparison demonstrates the limitations of the dTWA in describing the behaviour at intermediate and late times after a sudden quench from large transverse fields into the quantum critical regime.

For the quench protocol we consider here, spin-spin correlations of the TFIM reach stationarity already at relatively short times after the quench and are characterised by two correlation lengths \cite{Karl2017a}. 
The correlation length characterising the short-distance correlations is related to a generalised Gibbs ensemble (GGE).
We show that it is captured well by the dTWA simulations, except for regimes where stationarity is reached only at late times.
In contrast, the second correlation length, which characterises weak long-distance correlations, is not reproduced correctly in our semi-classical simulations.

\section{Transverse-Field Ising Model}
\begin{figure*}
  \includegraphics[width=\linewidth]{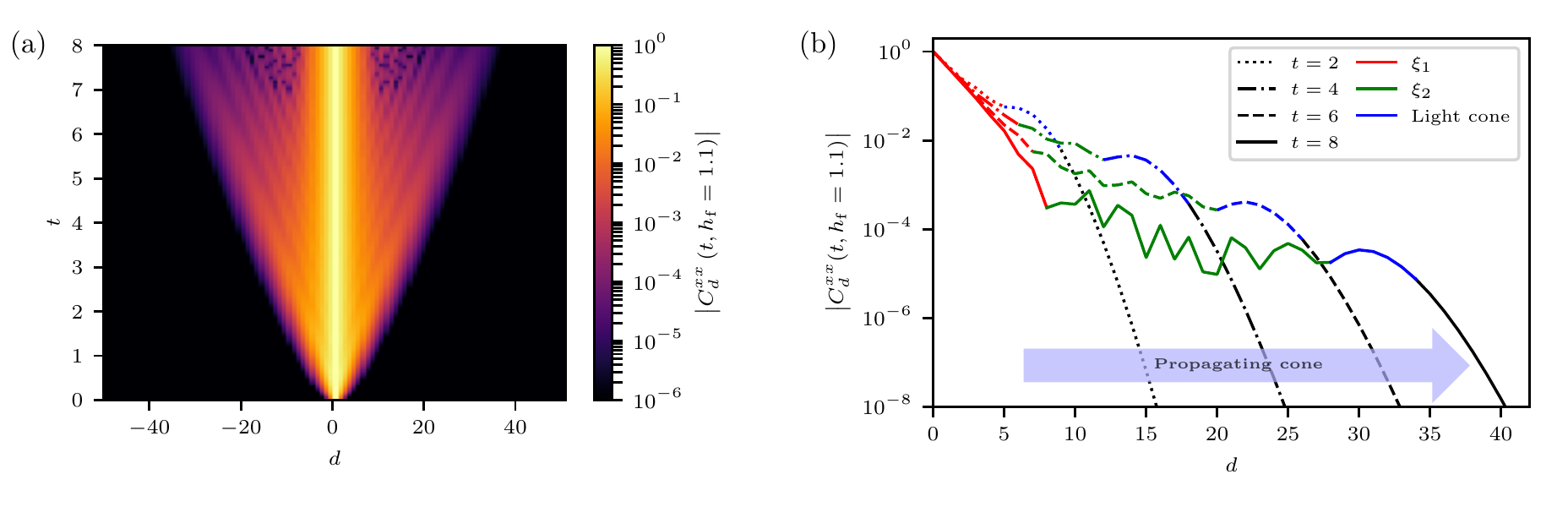}
  \caption{(a) Colorplot of the exact correlation function $C^{xx}_d=\left<\hat{\sigma}^x_0\hat{\sigma}^x_d\right>$ of the TFIM as a function of distance $d$ and time $t$ after a sudden quench from the ground state at $h_{\mathrm{i}}=1000$ to $h_{\mathrm{f}}=1.1$. The `light cone' can be observed to propagate linearly through the chain with time.
(b) Exact correlation function at times $t=2,4,6,8$ after the quench, corresponding to horizontal cuts through the right half of (a).
The colours denote different regimes, labelling the short-distance exponential fall-off $\sim\exp(-d/\xi_{1})$ (red), the intermediate-range oscillatory behaviour with exponential envelope (green) characterised by $\xi_{2}$, and the regime outside the `light cone' where the correlations fall off approximately proportional to a Gaussian (black).}
  \label{fig:Figure1}
\end{figure*}
The Hamiltonian of the TFIM reads
\begin{align}
  \hat{H}=-J\sum_{i=1}^N\hat{\sigma}^x_{i}\hat{\sigma}^x_{\left(i+1\right)\mathrm{mod}N}-h\sum_{i=1}^N\hat{\sigma}^z_i\,.
  \label{eq:Hamiltonian}
\end{align}
It describes a chain of $N$ quantum spins re\-pre\-sented by the Pauli matrices $\hat{\sigma}^j_{i}$, $i=1,\dots,N$, $j=x,y,z$, with nearest-neighbour interactions between the $x$-components of the spin and a field of strength $h$ applied in the $z$-di\-rection.
We assume periodic boundary conditions with an even number $N$ of sites. 
The model is integrable, and analytic solutions are obtained through Jordan-Wigner fermionisation \cite{Pfeuty1970a,Calabrese2012a,Calabrese2012b}. 
The TFIM possesses a quantum critical point at $h_{\mathrm{c}}=J$ and vanishing temperature, separating a ferromagnetic phase ($h<J$) from a paramagnetic phase ($h>J$), with the order parameter given by the magnetisation $\left<\hat{\sigma}^z\right>$ \cite{Sachdev1997,Sachdev2011}. Without loss of generality we set $J=1$ in the following.

We consider sudden quenches starting from a fully $z$-polarised initial state. 
For this, we prepare the system in the ground state of the TFIM Hamiltonian with a very large initial transverse field $h_{\mathrm{i}}$. At $t=0$ the transverse field is then switched instantaneously to its final value $h_{\mathrm{f}}>1$, which is chosen to be in the paramagnetic regime.

Analysing the exact solution of the TFIM, one finds that the correlation function \begin{equation}\label{e:Cxx}
C^{xx}_d\left(t\right)=\left<\hat{\sigma}^x_0\left(t\right)\hat{\sigma}^x_d\left(t\right)\right>
\end{equation}
reaches stationarity, at short distances $d$, already at fairly short times after the quench \cite{Karl2017a}. 
Before stationarity is reached, correlations are nonvanishing only inside a `light cone', as is illustrated in \Fig{Figure1}:
While, at short distances, the correlations fall off exponentially and oscillate, at the largest distances they approximately follow a Gaussian.
The scale where this Gaussian fall-off starts (marked in blue in  panel (b)), propagates out roughly linearly in time, defining a `light cone' in time $t$ and distance $d$, with velocity $v_{\mathrm{lc}}\simeq2$ for quenches within the paramagnetic regime, as can be observed in \Fig{Figure1} (a).
The velocity $v_{\mathrm{lc}}$ of the cone is given by the maximum group velocity as derived in the appendix, giving the propagation velocity of the elementary excitations. The maximum value is reached for the maximum occupation number in the fermion distribution and its stationarity leads to a light cone-like spreading of correlations.

Inside the cone, correlations change considerably in time, but the propagation of, e.g., the oscillating regime (marked in green in \Fig{Figure1} (b)) is much slower.
In the appendix, we summarise approximate analytical expressions for the correlations, as derived in  \cite{Calabrese2012a,Karl2017a}.

The (envelope of the) correlation function shows exponential fall-off with relative distance $d$ (red and green regions in \Fig{Figure1} (b)) \cite{Calabrese2012a,Karl2017a}.
After quenches into the vicinity of the quantum critical point, the short-distance behaviour is well described by a pure exponential with a correlation length $\xi_{1}$ that depends only on the final magnetic field $h_{\mathrm{f}}$ and not on time \cite{Karl2017a}.
At larger relative distances $d$, the correlations show oscillations, with an exponentially decreasing envelope, characterised by a second correlation length $\xi_{2}$ (green regime in \Fig{Figure1} (b)).
This behaviour dominates further away from the quantum critical point where the red regime shrinks to near zero. 

The stationary short-distance correlation length reads
\begin{equation}
  \label{eq:xi1hh0}
  \xi_{1}=\left[\mathrm{ln} \left(\frac{2h_\mathrm{i}h_\mathrm{f}}{h_\mathrm{i}+h_\mathrm{f}} \right)\right]^{-1}\,
\end{equation}
and is determined by a GGE, which is a generalisation of the canonical ensemble describing systems in thermal equilibrium.
The GGE is derived by translating the statistical entropy into the von-Neumann entropy of a density matrix \cite{Jaynes1957a,Jaynes1957b}. 
As discussed in \cite{Karl2017a}, $\xi_{1}$ can be extracted already at short times after the quench and for short chain lengths.

\section{Discrete Truncated Wigner Approximation}
\begin{figure}
  \includegraphics[width=\linewidth]{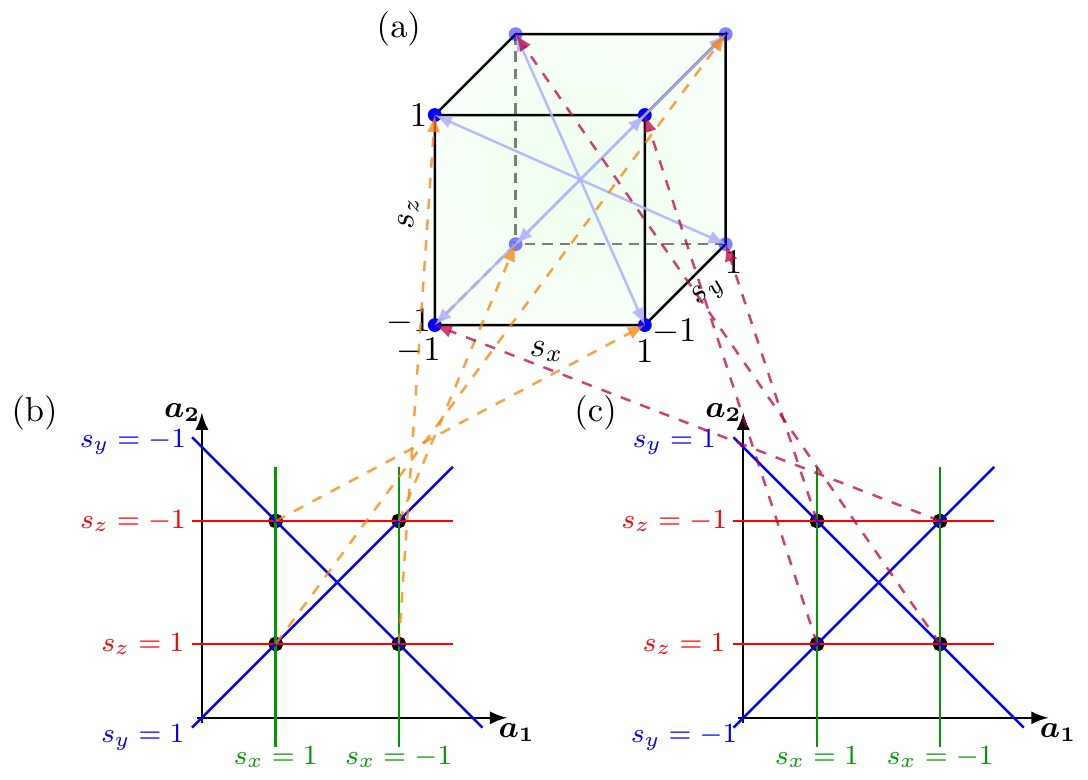}
  \caption{Sketch of the discrete phase space setup. 
  Each spin-$1/2$ particle can be in one of eight possible discrete spin states given by the configurations of each component being up or down, $s_{\alpha}=\pm1$. 
  These eight configurations are represented by the eight corners of the cube sketched in panel (a). 
  Panels (b) and (c) show two phase-space representations created using phase-point operators $\hat{A}_{\boldsymbol{\alpha}}$ [(b)] and $\hat{A}^{'}_{\boldsymbol{\alpha}}$ [(c)]. 
  Each phase-space representation is spanned by the variables $\boldsymbol{\alpha}=\left(a_1,a_2\right)$, $a_1,a_2\in\left\{0,1\right\}$. 
  The coloured lines denote sets of paired phase-space points $\boldsymbol{\alpha}$, where each set is associated with one spin component and each line connects the points representing either the $+1$ or the $-1$ state of the associated component. 
  Since the phase-point operators map one spin configuration onto each point in phase space, the association is defined by the phase-point operators and differs between the two phase-space representations in panels (b) and (c), representing four different spin states each. 
  The dashed arrows link the spin configurations in the cube in panel (a) with the associated phase-space points.
  Hence, by combining the two phase-space representations (b) and (c), all spin states are captured.}
  \label{fig:Figure2}
\end{figure}
The dTWA allows semi-classical simulations for the dynamics of discrete quantum systems, such as the spin-$1/2$ TFIM we consider here \cite{Schachenmayer2015,Schachenmayer2015b,Pucci2016}. 
The method is based on a discrete version of the Wigner representation of quantum mechanics, where states are described by quasi-probability distributions on phase space.
For spin-$1/2$ systems, this discrete quantum phase space consists of four points on which the state is defined by a two-dimensional vector over a real-valued finite field, $\boldsymbol{\alpha}=\left(a_1,a_2\right)$ with $a_1,a_2\in\left\{0,1\right\}$. 

Operators $\hat{\Omega}$ on Hilbert space can be mapped to Weyl symbols $\Omega_W$, which act on quasi-probability distributions on phase space. The mapping onto the Weyl symbols,
\begin{align}
  \Omega_W\left(\boldsymbol{\alpha}\right)=\frac{1}{M}\Tr{\hat{\Omega}\hat{A}_{\boldsymbol{\alpha}}},
  \label{eq:Weylsymbol}
\end{align}
is defined by phase-point operators $\hat{A}_{\boldsymbol{\alpha}}$, which map each basis state of the Hilbert space onto one point in phase space \cite{Wootters1987}. The choice of these phase-point operators is not unique.

One possible choice of phase-point operators is illustrated in \Fig{Figure2}(b) and given by
\begin{align}\label{e:rep1}
  \hat{A}_{\boldsymbol{\alpha}}=\frac{1}{2}\left[\left(-1\right)^{a_1}\hat{\sigma}^x+\left(-1\right)^{a_1+a_2}\hat{\sigma}^y+\left(-1\right)\hat{\sigma}^z+\hat{\mathbb{1}}\right],
\end{align}
which maps a spin configuration $\{s_x,s_y,s_z\}$ onto the four points in phase space in the following way:
\begin{align*}
  \left(s_x=1,s_y=1,s_z=1\right)&\rightarrow\left(a_1=0,a_2=0\right),\\
  \left(s_x=-1, s_y=-1, s_z=1\right)&\rightarrow\left(a_1=1,a_2=0\right),\\
  \left(s_x=1,s_y=-1,s_z=-1\right)&\rightarrow\left(a_1=0,a_2=1\right),\\
  \left(s_x=-1,s_y=1,s_z=-1\right)&\rightarrow\left(a_1=1,a_2=1\right).
\end{align*}

As the phase space consists of only four points, it only represents four of the eight existing spin configurations, where each component can be $+1$ or $-1$, corresponding to the eight corners of the cube in \Fig{Figure2}(a). 
By unitary transformations, additional phase-point operators $\hat{A}^{'}_{\boldsymbol{\alpha}}=\hat{U}\hat{A}_{\boldsymbol{\alpha}}\hat{U}^{\dagger}$ can be derived, resulting in a different phase-space representation. 
One possible result of this transformation is shown in \Fig{Figure2}(c), giving the phase-point operators
\begin{align}\label{e:rep2}
  \hat{A}^{'}_{\boldsymbol{\alpha}}=\frac{1}{2}\left[\left(-1\right)^{a_1}\hat{\sigma}^x-\left(-1\right)^{a_1+a_2}\hat{\sigma}^y+\left(-1\right)\hat{\sigma}^z+\hat{\mathbb{1}}\right]
\end{align}
which encodes the mapping
\begin{align*}
  \left(s_x=1,s_y=-1,s_z=1\right)&\rightarrow\left(a_1=0,a_2=0\right),\\
  \left(s_x=-1,s_y=1,s_z=1\right)&\rightarrow\left(a_1=1,a_2=0\right),\\
  \left(s_x=1,s_y=-1,s_z=-1\right)&\rightarrow\left(a_1=0,a_2=1\right),\\
  \left(s_x=-1,s_y=-1,s_z=-1\right)&\rightarrow\left(a_1=1,a_2=1\right).
\end{align*}
By combining the two phase space representations \eqref{e:rep1} and \eqref{e:rep2}, all eight spin configurations can be captured \cite{Schachenmayer2015b,Pucci2016}.
The Weyl symbol of the density operator $\hat{\rho}$ is called the Wigner function $W(\boldsymbol{\alpha})$. It is normalised to one but can have negative entries, which makes it a quasi-probability distribution on the discrete phase space. If a Wigner function happens to have only non-negative entries, it represents a proper probability distribution.

The key idea of the dTWA is to randomly sample spin configurations according to an initial, positive-definite probability distribution, and then to time-evolve each sample point \cite{Schachenmayer2015b,Pucci2016}.
The time evolution of each sample point can be calculated in the Schr\"odinger picture, where the Weyl symbols are time-independent, but the phase-point operators evolve in time, $\hat{A}_{\boldsymbol{\alpha}}(t)$. In general, this time evolution can not be calculated exactly and an approximation by truncation is necessary. For this we use a Bogoliubov-Born-Green-Kirkwood-Yvon (BBGKY) hierarchy, where reduced density operators are replaced by reduced phase-point operators, and truncate it at first or second order \cite{Pucci2016}. Truncation at first order gives the classical mean-field equations of motion \cite{Polkovnikov2010a}.

The dTWA procedure then involves sampling a large number $R$ of initial spin configurations, using the truncated time evolution for calculating trajectories, and calculating the desired observables as functions of time. 
By averaging the observables resulting from the individual samples, the dynamics of the system is simulated in an approximate, semi-classical manner \cite{Schachenmayer2015,Pucci2016}.

In this way, the time evolution of the correlation function $C^{xx}_d$ at a time $t$ after a sudden quench in the TFIM parameter $h$ can be simulated and a comparison with exact analytical solutions according to \cite{Pfeuty1970a,Calabrese2012a,Calabrese2012b} can be made. 
Here, we use this procedure for studying the capabilities and determining the limitations of the dTWA method for quenches within the paramagnetic phase.

\section{Results}
We simulate sudden quenches from fully $z$-polarised initial states (i.e., the ground state of the TFIM for very large initial transverse fields $h_\mathrm{i}$) to final transverse fields $h_{\mathrm{f}}>1$ and compare the results with the exact solution. 

We initially compare the two different sampling schemes, with the initial spin configurations $S_{4}$ obtained from the phase-space representation created by $\hat{A}_{\boldsymbol{\alpha}}$, see \Fig{Figure2}(b), and with $S_{8}$ obtained from the combination of the two phase space representations in \Fig{Figure2}(b) and (c), created by $\hat{A}_{\boldsymbol{\alpha}}$ and $\hat{A}^{'}_{\boldsymbol{\alpha}}$ as defined in \eqref{e:rep1} and \eqref{e:rep2}, respectively. 
Since $S_{8}$ yields more accurate results, we restrict ourselves to that sampling scheme in the remainder of this work. 

The initial state is numerically indistinguishable from a fully $z$-polarised state, with density matrix
\begin{align}
  \hat{\rho}=&\hat{\rho}_1\otimes\hat{\rho}_2\otimes\dots\otimes\hat{\rho}_N,\\
  \hat{\rho}_{i}=&\frac{1}{2}\left[\hat{\sigma}^z_i+\hat{\mathbb{1}}_i\right],
\end{align}
and one can show by plugging this into \Eq{Weylsymbol} that the resulting Wigner function is non-negative on both phase space representations considered here.

We consider the dynamics of the correlation function $C^{xx}_d$ in \eqref{e:Cxx} after a sudden quench and compare dTWA results with the exact solution obtained with the methods of \cite{Pfeuty1970a,Calabrese2012a,Calabrese2012b}. 
First we analyse how the accuracy of dTWA depends on time, in order to assess the timescales on which the approximation is valid.

\begin{figure}
  \includegraphics[width=\linewidth]{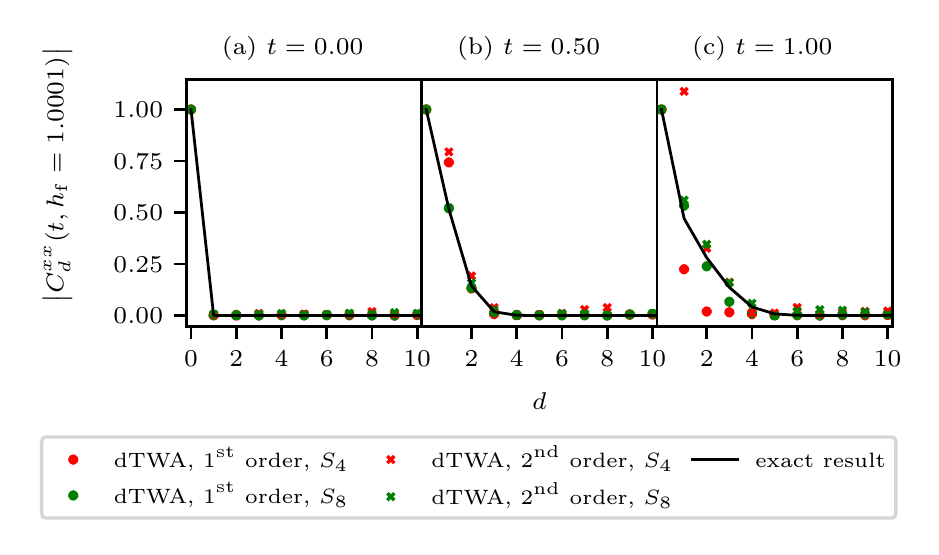}
  \caption{
  Correlation function $C^{xx}_d\left(t,h_{\mathrm{f}}\right)$ at times $t=0$, $0.5$, $1$ after a sudden quench to $h_{\mathrm{f}}=1.0001$. 
  First-order dTWA results ($R=10000$ samples) and second-order ($R=1000$), sampling from both, $S_4$ and $S_8$, are compared to the exact solution. A linear ordinate places the focus on the short-distance behaviour.}
  \label{fig:Figure3}
\end{figure}

\Fig{Figure3} shows the absolute value of the correlation function at different times $t$ after a quench to $h_{\mathrm{f}}=1.0001$, very close to the quantum critical point.
Since $\hat{\sigma}^x_i\hat{\sigma}^x_i=\hat{\mathbb{1}}$, the correlation function at relative distance $d=0$ is given by the normalisation of the state \cite{Pucci2016}.
By construction, the initial state at $t=0$ is reproduced exactly up to statistical fluctuations due to the finite sample size. 
At times $t>0$, systematic errors stemming from the quasi-classical approximation of the equations of motion start to appear in addition to the statistical fluctuations. 
These systematic errors are smaller in the second-order truncation, as one would expect. 
The second-order dTWA with sampling from $S_8$ matches the exact solution with good accuracy, even at times $t\approx1$ where a first-order truncation shows already substantial deviations from the exact result. 
At even later times, however, the second-order truncation suffers from instabilities of the equations of motion, as they have been discussed in \cite{Ryzhov2000,Berges2004}, which leads to unphysical results. 
It is for this reason that, for the late-time dynamics we want to analyse, we will mostly be using a first-order truncation scheme in the following.

\begin{figure*}
  \includegraphics[width=\linewidth]{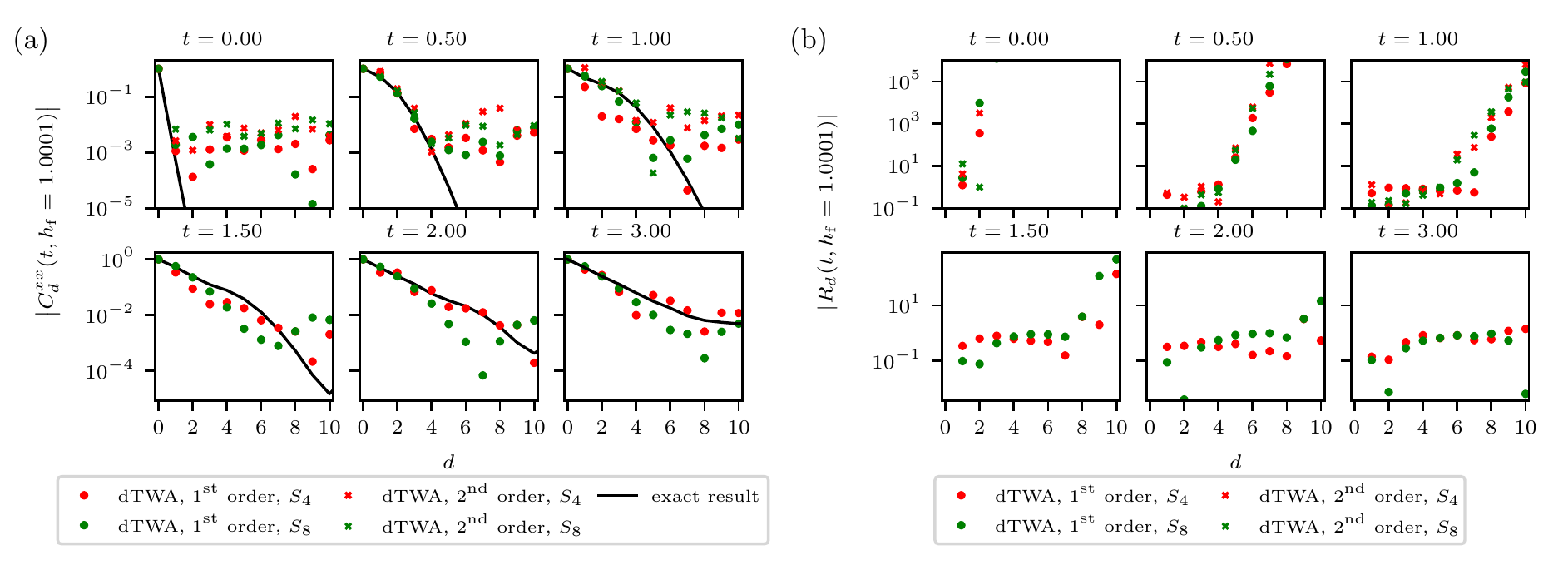}
  \caption{
  (a) 
  Correlation function $C^{xx}_d\left(t,h_{\mathrm{f}}\right)$ at different times after a sudden quench to $h_{\mathrm{f}}=1.0001$, on a semi-logarithmic scale. 
  The dTWA simulations of first and second order using sampling schemes from the sets $S_4$ and $S_8$ are compared to the exact solution. The second-order dTWA results are unstable at late times and are not shown.
  (b)
  Residuals $R_d\left(t,h_{\mathrm{f}}\right)$ obtained by subtracting the exact correlation functions from the dTWA data for  $C^{xx}_d\left(t,h_{\mathrm{f}}\right)$, 
normalised to the exact value of the correlations, 
for the same parameters and approximations as shown in the left panels.  
  }
  \label{fig:Figure4}
\end{figure*}
Due to the exponential fall-off, the behaviour at large distances is difficult to see on the linear scale of \Fig{Figure3}. 
To focus more on the long-distance tails, \Fig{Figure4} shows the correlation functions on a logarithmic scale, as well as the residuals obtained by subtracting the exact solution from the dTWA results and dividing the result by the exact solution. 
Here, also longer times up to $t=3$ are considered. 
As becomes evident from these graphs, the relative deviations are of the same size for all times. The divergence at large distances and short times results from the extremely small exact values reached due to the exponential decay.

\begin{figure}
  \includegraphics[width=\linewidth]{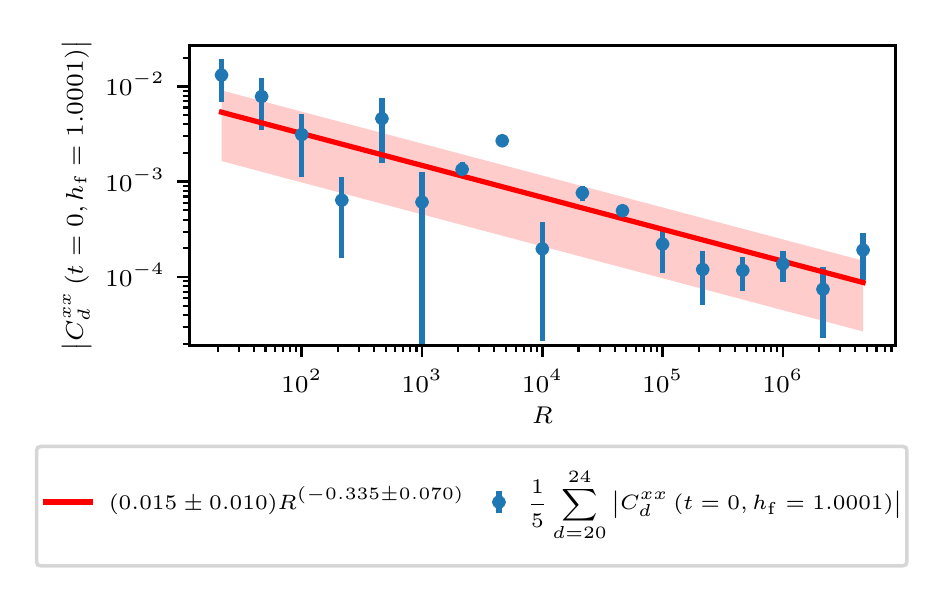}
  \caption{Absolute values of the initial correlation function $C^{xx}_d\left(t=0,h_{\mathrm{f}}=1.0001\right)$ with $h_{\mathrm{i}}=1000$ averaged over large relative distances $d=20,\dots,25$, deep in the plateau reached in the initial state.
The data is shown for a chain with $N=50$ sites, as a function of sampled runs $R$ in the dTWA, sampling from the set $S_8$. Error bars show the standard deviation of the distribution of $C_d^{xx}$ ($d=20,\dots,24$). At these relative distances, the simulations reach a plateau caused by limited numerical precision due to finite sampling. With increasing $R$, the precision can be increased and a function $f\left(R\right)=aR^b$ has been fitted to describe the dependence of the precision on $R$, giving $a=0.015\pm0.01$, $b=-0.335\pm0.07$, which is also plotted with the fitting error denoted by the shaded region.}
  \label{fig:Figure5}
\end{figure}
For the initial state ($t=0$ shown in \Fig{Figure4}(a), top left panel), where we know that only statistical errors are present, we can read off the size of that error from the value of the plateau visible in the plot.
We study the size of the error as a function of the number of samples $R$ in the first-order dTWA.
The result is shown in \Fig{Figure5}, where we depict, for different $R$, the correlation function $C^{xx}_d\left(t=0,h_{\mathrm{f}}=1.0001\right)$ at $t=0$ averaged over large relative distances $d=20,\dots,24$ in a spin chain with $N=50$ sites, i.e., in a regime where the plateau is reached. 
The graph demonstrates that the plateau is of statistical origin, and that it can be reduced by increasing $R$. 
\begin{figure}
  \includegraphics[width=\linewidth]{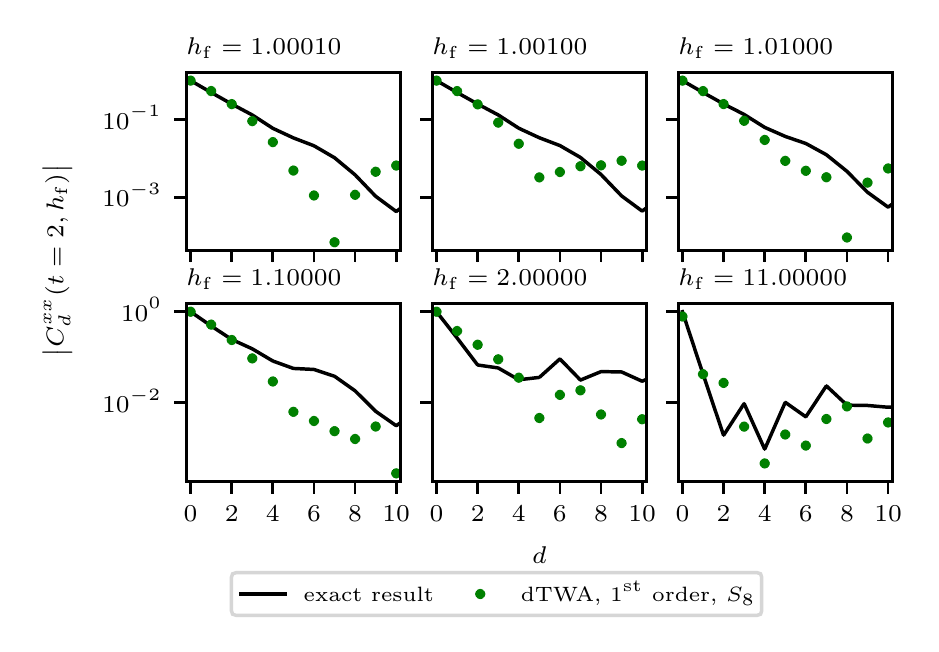}
  \caption{ 
  Correlation function $C^{xx}_d\left(t,h_{\mathrm{f}}\right)$ at time $t=2$ after a sudden quench to different final fields. The first-order dTWA simulations using sampling set $S_8$ are compared with the exact solutions.
}
  \label{fig:Figure6}
\end{figure}
\begin{figure*}
  \includegraphics[width=\linewidth]{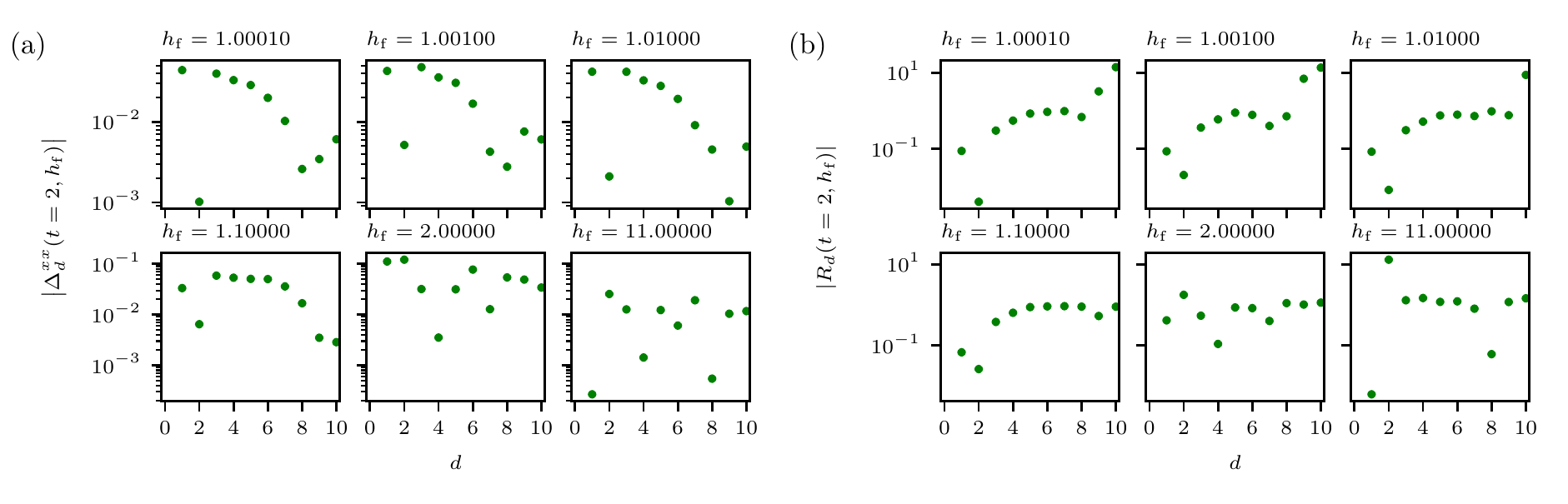}
  \caption{
  (a)  
  Absolute errors $\Delta^{xx}_d\left(t=2,h_{\mathrm{f}}\right)$ obtained by subtracting the exact correlation functions from the  simulated $C^{xx}_d\left(t,h_{\mathrm{f}}\right)$ for the data shown in \Fig{Figure6}.
  (b) 
  Residuals $R_d\left(t=2,h_{\mathrm{f}}\right)$ obtained by dividing the errors $\Delta^{xx}_d\left(t=2,h_{\mathrm{f}}\right)$ by the exact value of the correlations $C^{xx}_d\left(t=2,h_{\mathrm{f}}\right)$.
  Errors are expected to diminish with growing sample size, cf.~\Fig{Figure5}.
}
  \label{fig:Figure7}
\end{figure*}

By fitting a function $f\left(R\right)=aR^b$ to the data, we find the parameters $a=0.015\pm0.01$, $b=-0.33\pm0.07$ to describe the dependence of the statistical error on $R$. 
Hence, we need to choose $R\approx10^{6}$ to capture correlation functions of size $C^{xx}_d\approx10^{-4}$. 
This leads to excessively large computation times, such that $R$ can not be increased far enough to reach the precision needed for capturing the exponential fall-off known for the exact solution at large relative distances $d$. 
In the following, we choose $N=20$ spin sites and impose periodic boundary conditions, which represents a compromise between independence of finite-size effects and computation time. For all first-order simulations we choose $R=10000$ and for second-order simulations we choose $R=1000$.

Next, we investigate how the accuracy of the method depends on the field $h_\mathrm{f}$ after the quench. 
\Fig{Figure6} shows the correlation function at a fixed time $t=2$ after quenches to different values of $h_{\mathrm{f}}$, using the first-order equations applied to sampling sets $S_{8}$. For the same setting and parameter values, \Fig{Figure7} shows the absolute errors (left panels) and the corresponding residuals (right panels). 
While the relative error is generally found to be on the order of and below $0.1$ only for the first two points with $d>0$, it grows above ten percent in the further tail.
The absolute error is below $0.1$ for all data points.
The dTWA appears to capture the plateau at large $d$ for quenches further away from the critical point (last panel of \Fig{Figure6}).
The relative errors, in these cases, are, however, on the order of $1$ as for quenches closer to criticality.
We expect that an increase in sample size $R$ may help to reach a better accuracy for larger distances $d$ at least further away from the critical point, which for our current computational resources is out of reach.

\begin{figure}
  \includegraphics[width=\linewidth]{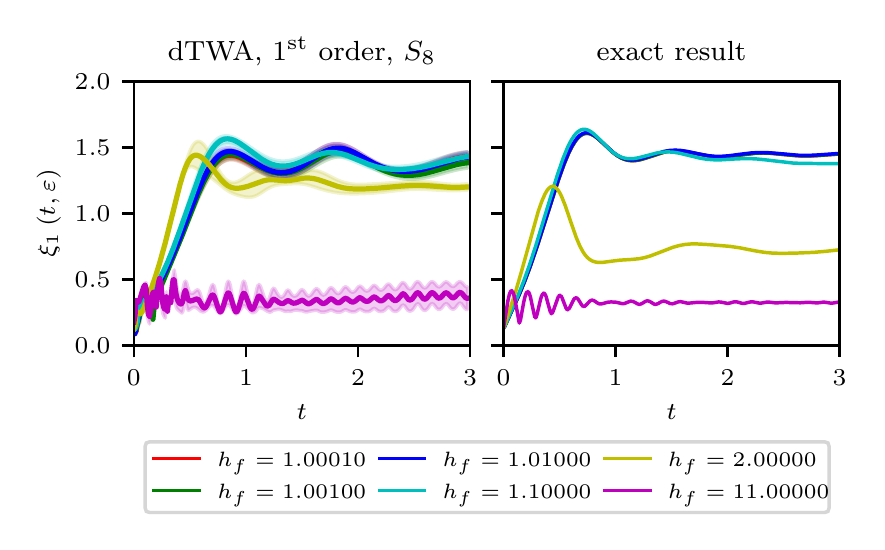}
  \caption{
  Time evolution of the correlation length $\xi_{1}$ after a sudden quench from $h_{\mathrm{i}}=1000$ to different final transverse fields $h_\mathrm{f}=1+\varepsilon$. 
  The first-order dTWA simulations using sampling from the sets $S_8$ are compared to the exact solutions. 
  Shaded regions denote the fitting uncertainty. 
  At short times, when the correlation function agrees with the exact solution, also the correlation length captures the exact behaviour well.
  }
  \label{fig:Figure8}
\end{figure}
To further analyse the performance of the dTWA at short distances, we extract $\xi_1$ from the correlation function by fitting an exponential function to the short-distance decay, $C^{xx}_d\propto\mathrm{exp}\left(-d/\xi_{1}\right)$ ($d<3$). We consider the time evolution of $\xi_1\left(t,\varepsilon\right)$ after quenches to different distances $\varepsilon$ from the quantum critical point at $h_{\mathrm{c}}=1$,
\begin{align}
\varepsilon=\frac{h_{\mathrm{f}}-h_{\mathrm{c}}}{h_{\mathrm{c}}}=h_{\mathrm{f}}-1,
\end{align}
and compare dTWA data with exact values for $\xi_{1}$ as functions of $\varepsilon$, as shown in \Fig{Figure8}.

Since the correlation length is extracted from the short-distance correlation function, it captures the short-time behaviour well, as at these times the correlation function also describes the exact solution with good accuracy, see \Fig{Figure4}. Close to and after the first maximum of the damped oscillations the data deviates from the exact correlation length. 
While the oscillations are not captured well, the simulations saturate, however, near the exact values in the long-time limit, at least for sufficiently small deviations of $h_\mathrm{f}$ from the critical point

At long times, the exact correlation length converges to a stationary value that is determined by the diagonal elements of the density matrix characterising the initial state after the quench. The corresponding diagonal ensemble matrix takes the form of a GGE and is set by the Bogoliubov fermion occupation numbers. The occupations are stationary after the sudden quench and are hence defined by the initial and final transverse field \cite{Karl2017a}.
In the limit of an infinitely large initial transverse field, the asymptotic, i.e., late-time correlation length resulting from the GGE is well described by the universal crossover function
\begin{align}
  \xi_{\mathrm{GGE}}\left(\varepsilon\right)=\frac{1}{\ln{2\varepsilon+2}},
\end{align}
which depends on the distance $\varepsilon$ from the quantum critical point \cite{Calabrese2012a,Karl2017a}.
\begin{figure}
  \includegraphics[width=\linewidth]{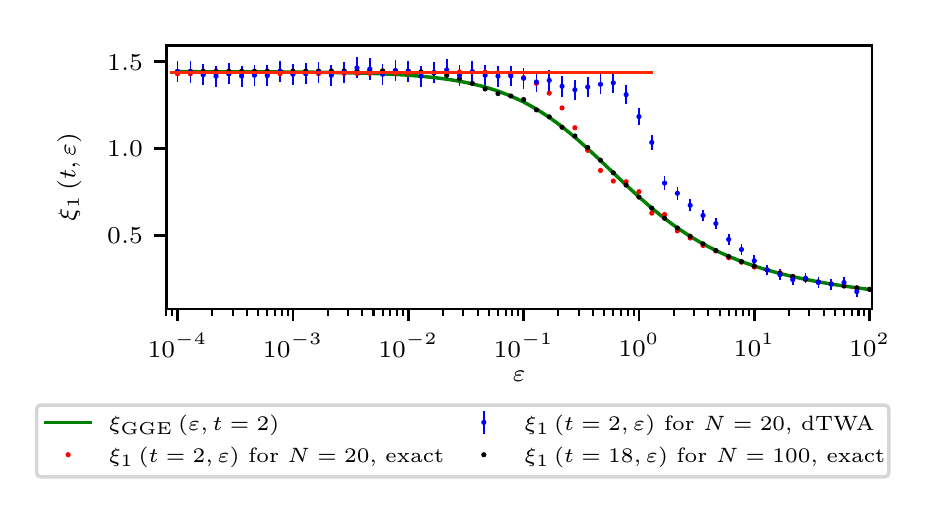}
  \caption{
  Correlation length $\xi_{1}$ at a fixed time $t=2$ after a sudden quench from $h_{\mathrm{i}}=1000$ as a function of the final relative distance $\varepsilon=h_{\mathrm{f}}-1$ from the quantum critical point. 
  First-order dTWA simulations using sampling from the configuration sets $S_8$ are compared with the exact solution. The function $\xi_{\mathrm{GGE}}(\varepsilon)$ describing the correlation length obtained for a GGE is shown as a green line and describes the stationary behaviour of the exact solution at late times. The numerically exact data at time $t=18$ (black points; using $N=100$ sites to avoid finite-size effects) is shown for verification. Errorbars denote the statistical uncertainty.
}
  \label{fig:Figure9}
\end{figure}
\Fig{Figure9} shows the correlation length at a fixed time $t=2$ after a sudden quench to $\varepsilon$, comparing the dTWA simulations with the exact analytical result as well as $\xi_{\mathrm{GGE}}(\varepsilon)$. 
While the exact solution as well as the dTWA data converge to $\xi_{\mathrm{GGE}}(\varepsilon)$ at large and small $\varepsilon$, at intermediate distances from the critical point, stationarity is not yet reached at this time. In this regime, the dTWA simulations show deviations from the exact solution, which we have found to remain even for later times, when the exact solution can everywhere be described by $\xi_{\mathrm{GGE}}\left(\varepsilon\right)$.
At the quantum critical point, the correlation length does not diverge, but saturates at a finite value.
\begin{figure}
  \includegraphics[width=\linewidth]{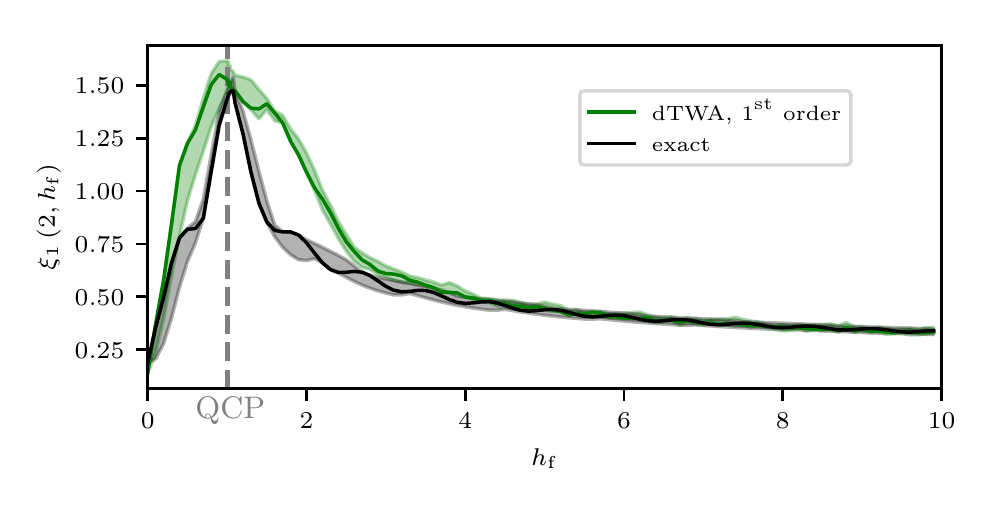}
  \caption{
  Correlation length $\xi_{1}$ at a fixed time $t=2$ after a sudden quench from $h_{\mathrm{i}}=1000$ within the paramagnetic and into the ferromagnetic regime as a function of the final transverse field $h_{\mathrm{f}}$. For $h_{\mathrm{f}}>1$ the data is the same as in \Fig{Figure8}.
  First-order dTWA simulations using the sampling set $S_8$ are shown in comparison with the exact solution. The shaded regions indicate the amplitude of the temporal oscillations remaining at the latest evolution times computed.
  The lower and upper limits of the shaded region correspond to the minimum and maximum value, respectively, within the interval from $t=1.3$ to $t=2$. The dashed vertical line indicates the field where the quantum phase transition occurs in the ground state.}
  \label{fig:Figure10}
\end{figure}

Finally, we also study quenches across the quantum critical point into the ferromagnetic regime. \Fig{Figure10} shows the correlation length at a fixed time $t=2$ after a quench to final fields between $h_{\mathrm{f}}=0$ and $h_{\mathrm{f}}=10$, on a linear scale for $h_\mathrm{f}$.
Close to $h_{\mathrm{f}}=0$ the dTWA captures the exact solution very well, as in this regime the transverse field is a perturbation to the classical Ising model and the dynamics can be described via a low-order cumulant expansion \cite{Schmitt2017a}.
Around the quantum critical point, deviations appear in both, the ferromagnetic and the paramagnetic regime, while further away from the quantum critical point in the paramagnetic regime the dTWA again captures the exact solution very well. This exhibits again the relative limitations of the dTWA in the vicinity of the quantum critical point, see also \cite{Czischek2018}.

\section{Conclusions}
By applying the discrete truncated Wigner approximation (dTWA) to the transverse-field Ising model and considering dynamics after sudden quenches to different distances from the quantum critical point, we benchmarked the dTWA against exact results and discuss the performance of the simulation method in different regimes. We focus in particular on quenches close to the quantum critical point where entanglement grows indefinitely in time and MPS-based simulation methods are struggling. For this very reason, this is also the regime where quantum simulation may prove particularly useful. In the close vicinity of the quantum critical point, we find the semi-classical dTWA to give results close to the exact solution, while only at intermediate distances $\varepsilon$ we find deviations, which disappear again when considering larger transverse fields $h_{\mathrm{f}}\gg 1$.

Truncating the semi-classical equations of motion at second order, we find that the dTWA approximates the exact correlation functions with good accuracy for short times, much better than in a first-order approximation. At later times, however, the second-order equations of motion suffer from instabilities, which lead to unphysical results. For this reason, in order to be able to analyse the stationary regime at later times, we mostly focused on dTWA with a first-order truncation scheme, where instabilities do not occur, see \Fig{Figure4}.

For the quenches within the paramagnetic phase that we studied, the dTWA captures the exponentially decaying correlation function at short spatial distances very well.
However, correlations are reproduced only down to a minimum value, due to statistical errors caused by the finite sample size. 
This limit can be shifted to smaller values by increasing the sample size, but, as we have shown in \Fig{Figure5}, it decreases only slowly with increasing the number of samples, so that one is restricted to regimes where larger correlations prevail.
The correlation length $\xi_1$ which determines the exponential fall-off at short spatial distances, is well captured by the dTWA simulations, see Figs.~\fig{Figure8}, \fig{Figure9}, and \fig{Figure10}.

At larger distances $\varepsilon$ from the quantum critical point, a secondary oscillatory behaviour with exponentially decaying envelope is known to occur \cite{Karl2017a}.
Corresponding results could not be detected within the dTWA approach here, likely due to insufficient sample sizes, as shown in Figs.~\fig{Figure6} and \fig{Figure7}. 
The secondary fall-off, however, may also be impossible to be described within a semi-classical approach but rather require a full quantum calculation or simulation. 

For quenches to intermediate distances $\varepsilon$ from the quantum critical point, $0.1\lessapprox\varepsilon=h_{\mathrm{f}}-1\lessapprox10$, the late-time dTWA simulations give results for the correlation length $\xi_{1}$ which saturate at values deviating from the exact solution. 
In the regimes very close to criticality as well as further away from the critical point the late-time asymptotic correlation length is however captured very well in the simulations.

All in all, for the quenches in the transverse-field Ising model considered in this paper, second-order dTWA gives very accurate results for correlation functions at short times, before divergencies appear. First-order dTWA gives fair results on short to intermediate timescales, but has the great advantage that the method does not suffer from instabilities. The late-time asymptotics of the correlation function is correctly reproduced in first-order dTWA, except for quenches to intermediate distances from the quantum critical point. Correlations between far distant sites are challenging to simulate in first- and second-order dTWA alike, as statistical fluctuations due to the initial state sampling are larger than the small values of the correlation function at large distances.

Our results reveal that a good understanding of the validity and accuracy of dTWA simulations will prove beneficial for comparison and benchmarking of such classical simulations with quantum simulation results, in a regime that is challenging to access with other methods.

\section*{Acknowledgements} 
The authors thank J.~Halimeh, M.~Karl, A.~Pi{\~n}eiro Orioli, A.-M.~Rey, J. Schachenmayer, and C.-M.~Schmied for discussions and collaboration on the topics described here. 
This work was supported by the Horizon-2020 framework programme of the European Union (FET-Proactive, AQuS, No. 640800, ERC Ad\-vanced Grant EntangleGen, Project-ID 694561),  by Deutsche Forschungsgemeinschaft (SFB 1225 ISOQUANT), and by Heidelberg University (CQD, STRUCTURES).
M. K. acknowledges financial support from the National Research Foundation of South Africa through the Competitive Programme for Rated Researches.
T.G., M.O. and M.K. thank the Erwin Schr\"odinger International Institute for Mathematics and Physics, for hospitality and support.
\\


\begin{appendix}
\begin{center}
\end{center}
\setcounter{equation}{0}
\setcounter{table}{0}
\makeatletter

\section*{Appendix: Transverse-field Ising model: Analytical results}
We briefly summarise the approximate analytical behaviour of the correlations after the quench from $h_\mathrm{i}$ to $h_\mathrm{f}$ (see \cite{Calabrese2012a,Karl2017a} for details). Note that for the calculations in the main text we used exact results requiring a numerical diagonalisation step. 
For quenches within the paramagnetic phase, the correlations between the $x$-components of the spins are approximately given by
\begin{align}
  \nonumber
  &C^{xx}_d\left(t,h_{\mathrm{f}}\right)\simeq C_0\left(h_{\mathrm{i}},h_{\mathrm{f}}\right)e^{-d/\xi_1\left(h_{\mathrm{i}},h_{\mathrm{f}}\right)}+\left(h_{\mathrm{f}}^2-1\right)^{1/4}\sqrt{4h_{\mathrm{f}}}\\
  \nonumber
  &\times\int_{-\pi}^{\pi}\frac{\mathrm{d}k}{\pi}\left[\frac{n_{\mathrm{BF}}\left(k\right)}{1-n_{\mathrm{BF}}\left(k\right)}\right]^{1/2}\frac{\mathrm{sin}\left[2\omega_{\mathrm{BF}}\left(k;h_{\mathrm{f}}\right)t-kd\right]}{\omega_{\mathrm{BF}}\left(k;h_{\mathrm{f}}\right)}\\
  \nonumber 
  &\times\mathrm{exp}\left(\int_{0}^{\pi}\frac{\mathrm{d}k}{\pi}\mathrm{ln}\left|1-2n_{\mathrm{BF}}\left(k\right)\right|\right.\\
  &\left.\vphantom{\int_{0}^{\pi}\frac{\mathrm{d}k}{\pi}}\times\left\{d+\Theta\left[d-2v_{\mathrm{BF}}\left(k;h_{\mathrm{f}}\right)t\right]\left[2v_{\mathrm{BF}}\left(k;h_{\mathrm{f}}\right)t-d\right]\right\}\right),
  \label{eq:Corr}
\end{align}
where $\Theta$ denotes the Heaviside step function. This expression depends on the amplitude
\begin{align}
  C_0\left(h_{\mathrm{i}},h_{\mathrm{f}}\right)=\left[\frac{\left(h_{\mathrm{i}}-h_{\mathrm{f}}\right)h_{\mathrm{f}}\sqrt{h_{\mathrm{i}}^2-1}}{\left(h_{\mathrm{i}}+h_{\mathrm{f}}\right)\left(h_{\mathrm{f}}h_{\mathrm{i}}-1\right)}\right]^{1/2},
\end{align}
for $h_{\mathrm{i}}>h_{\mathrm{f}}$, and on the inverse correlation length
\begin{align}
  \nonumber
  \xi^{-1}_1\left(h_{\mathrm{i}},h_{\mathrm{f}}\right)
  &= \Theta\left(h_{\mathrm{f}}-1\right)\Theta\left(h_{\mathrm{i}}-1\right)\ln{\mathrm{min}\left\{h_{\mathrm{i}},h_{1}\right\}}\\
  &\ \ -\ \frac{1}{2\pi}\int_{-\pi}^{\pi}\mathrm{d}k\mathrm{ln}\left|1-2n_{\mathrm{BF}}\left(k\right)\right|.
\end{align}
Here $h_1$ is defined as
\begin{align}
  h_1=\frac{1+h_{\mathrm{f}}h_{\mathrm{i}}+\sqrt{\left(h_{\mathrm{f}}^2-1\right)\left(h_{\mathrm{i}}^2-1\right)}}{h_{\mathrm{f}}+h_{\mathrm{i}}}.
\end{align}
In \Eq{Corr},
\begin{equation}
  n_{\mathrm{BF}}\left(k;h_{\mathrm{i}},h_{\mathrm{f}}\right)
  = \frac{1}{2}
  -2\frac{h_{\mathrm{i}}h_{\mathrm{f}}+1-\left(h_{\mathrm{i}}+h_{\mathrm{f}}\right)\cos{k}}
      {\omega_{\mathrm{BF}}\left(k;h_{\mathrm{f}}\right)\omega_{\mathrm{BF}}\left(k;h_{\mathrm{i}}\right)}
\end{equation}
are the mode occupation numbers of the Bogoliubov fermions diagonalising the Hamiltonian \Eq{Hamiltonian} after the quench, and
\begin{equation}
  \omega_{\mathrm{BF}}\left(k;h\right)= 2\sqrt{h^2+1-2h\cos{k}}
\end{equation}
are the corresponding mode frequencies. From the latter, the group velocity
\begin{align}
  v_{\mathrm{BF}}\left(k;h\right)
  =\frac{\mathrm{d}\omega_{\mathrm{BF}}\left(k;h\right)}{\mathrm{d}k}
\end{align}
can be derived.

\end{appendix}

\end{document}